\documentclass[
 reprint,
 amsmath,amssymb,
 aps,
 footinbib,
 superscriptaddress,
 pra,
]{revtex4-2}

\usepackage{placeins}
\usepackage{lipsum}
\usepackage{graphicx}
\usepackage{epstopdf}

\usepackage{dcolumn}
\usepackage{bm}
\usepackage[colorlinks=true, citecolor={blue!80!black}, urlcolor={blue!50!black}, linkcolor = {blue!80!black}]{hyperref}
\usepackage[colorinlistoftodos]{todonotes}
\usepackage[capitalize]{cleveref}
\usepackage[english]{babel}
\usepackage{braket}
\usepackage[utf8]{inputenc}
\usepackage{physics}

\usepackage{siunitx}
\DeclareSIUnit{\belmilliwatt}{Bm}
\DeclareSIUnit{\dBm}{\deci\belmilliwatt}

\begin{document}

\preprint{APS/123-QED}

\title{Blueprint for all-to-all connected superconducting spin qubits}

\author{Marta Pita-Vidal}
\affiliation{QuTech and Kavli Institute of Nanoscience, Delft University of Technology, 2600 GA Delft, The Netherlands}

\author{Jaap J. Wesdorp}
\affiliation{QuTech and Kavli Institute of Nanoscience, Delft University of Technology, 2600 GA Delft, The Netherlands}

\author{Christian Kraglund Andersen}
\affiliation{QuTech and Kavli Institute of Nanoscience, Delft University of Technology, 2600 GA Delft, The Netherlands}

\date{\today}

\begin{abstract}

Andreev (or superconducting) spin qubits (ASQs) have recently emerged as a promising qubit platform that combines superconducting circuits with semiconductor spin degrees of freedom. 
While recent experiments have successfully coupled two ASQs, how to realize a scalable architecture for extending this coupling to multiple distant qubits remains an open question.
In this work, we resolve this challenge by introducing an architecture that achieves all-to-all connectivity between multiple remote ASQs.
Our approach enables selective connectivity between any qubit pair while maintaining all other qubit pairs uncoupled.
Furthermore, we demonstrate the feasibility of efficient readout using circuit quantum electrodynamics techniques and compare different readout configurations.
Our architecture shows promise both for gate-based quantum computing and for analog quantum simulation applications by offering higher qubit connectivity than alternative solid-state platforms.

\end{abstract}

\maketitle

To date, two of the most scalable solid-state qubit platforms are semiconducting spin qubits and superconducting qubits.
Recent experiments in these architectures have realized systems with tens and hundreds of qubits, respectively~\cite{Philips2022, Borsoi2023, Wang2024, Neyens2024, Kim2023, Shtanko2023, Arute2019, Google2023b, Google2023}, 
as well as two-qubit gates with fidelities above 99~\%~\cite{Mills2022, Xue2022, Noiri2022, Madzik2022, Wu2024, Tanttu2023, Sung2021, Ding2023, Sheldon2016,  Hong2020, Wang2024, Marxer2023}.
However, both platforms are currently limited to low qubit connectivity, often restricted to nearest neighbors in planar grids, and typically featuring at most three to five connections per qubit~\cite{Vandersypen2017, Li2018,  Tadokoro2021, Boter2022, Nguyen2022, Versluis2017, Arute2019, Chamberland2020, Google2023}. 
This sparse connectivity results in a considerable overhead in qubit count, both when it comes to error correction codes in gate-based quantum computation applications~\cite{Bravyi1998, Chamberland2020, Dennis2022, Bravyi2024} and to analog quantum simulations~\cite{Lechner2015, Baumer2023}.

An alternative platform to the aforementioned qubits are Andreev (or superconducting) spin qubits (ASQs)~\cite{Chtchelkatchev2003, Padurariu2010, Park2017, Hays2021, PitaVidal2023, PitaVidal2024, Spethmann2022, Hoffman2024}.
These qubits have their state encoded in the spin of a quasiparticle localized within a semiconducting quantum dot that is tunnel-coupled to two superconducting leads, thus forming a Josephson junction~\cite{Chtchelkatchev2003, Padurariu2010}.
Recent experimental realizations have explored systems with a single ASQ~\cite{Hays2021, PitaVidal2023} as well as the supercurrent-mediated coupling between two distant ASQs~\cite{PitaVidal2024}.
Yet, no experiments involving more than two ASQs have been reported. 
Architectures for coupling either adjacent Andreev qubits via wavefunction overlap~\cite{Spethmann2022} or two distant qubits via virtual photons \cite{Cheung2023} have been proposed.
However, these architectures enable the short-distance coupling of nearest-neighbour qubits in a planar layout plus a reduced number of long-distance links, sharing the same connectivity constraints as semiconducting and superconducting qubits. Thus, it remains an open question if the compact size of ASQs and their strong coupling to supercurrent can be combined to provide architectural improvements over more conventional solid-state platforms.

Here, we introduce an architecture that offers a solution to the connectivity challenge.
Our design allows for the coupling of multiple distant ASQs in a fully connected and scalable way.
As a particular example, we demonstrate that this architecture permits selective connectivity between any qubit pair within the system while maintaining all other qubit pairs uncoupled. 
Importantly, the strength of the coupling between two qubits in the system is independent of the distance between them.
Furthermore, we illustrate how this system can efficiently perform quantum simulations of highly connected Ising models with a reduced qubit count and a smaller footprint compared to alternative solid-state platforms.
The proposed architecture also facilitates sequential, individual, or joint qubit readout.
Finally, we outline an experimental protocol for systematically tuning up the system to its operational setpoint.

\section{Concept}\label{s:scalability_concept}

\begin{figure*}
    \centering
    \includegraphics{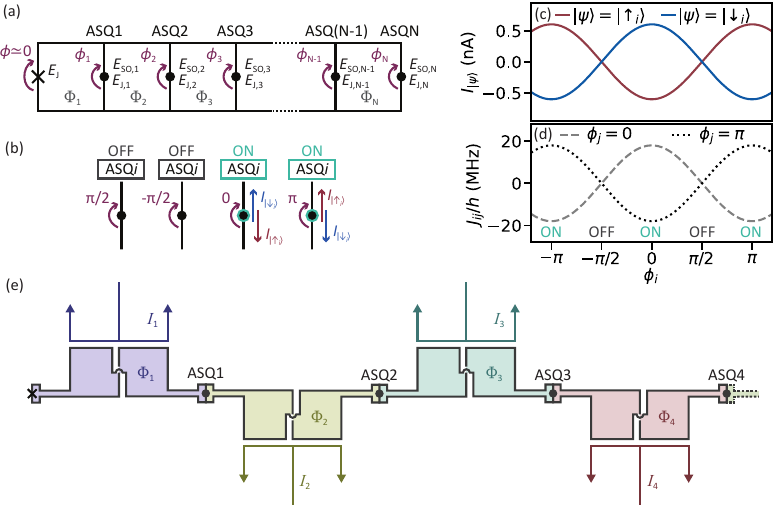} 
    \caption{\label{fig:scalability_concept}
    {\bf Scalability of superconducting spin qubits.} 
    (a) Circuit diagram of $N$ Andreev spin qubits connected in parallel to a coupling Josephson junction with Josephson energy $E_{\rm J}$, thus defining $N$ loops. 
    ASQ$i$ has spin-independent and spin-dependent Josephson energies $E_{{\rm J},i}$ and $E_{{\rm SO},i}$, respectively. 
    The magnetic flux through loop $i$ is denoted as $\Phi_i$.
    The phase drop across ASQ$i$ is denoted as $\phi_i$ and that across the coupling junction as $\phi$. 
    The system is operated in the regime $E_{\rm J} \gg E_{{\rm J},i}, E_{{\rm SO},i}$, which results in $\phi \simeq 0$.
    (b) Diagrams of four possible phase setpoints for an ASQ. 
    When $\phi_i=\pi/2,-\pi/2$ the spin-dependent component of the supercurrent vanishes and the qubit is labeled as OFF.
    When instead $\phi_i=0, \pi$ the spin-dependent component of the supercurrent is maximal and the qubit is labeled as ON. We use the notation of $I_{\ket{\uparrow_i}}=\bra{\uparrow_i}\mathcal{I}_i\ket{\uparrow_i}$ and $I_{\ket{\downarrow_i}}=\bra{\downarrow_i}\mathcal{I}_i\ket{\downarrow_i}$.
    (c) Supercurrent across ASQ$i$ versus the phase drop across it for its two basis states, $\ket{\uparrow_i}$ and $\ket{\downarrow_i}$. Here we assume $E_{{\rm SO}, i}/h = $~\SI{300}{MHz}, $E_{{\rm J}, i}/h = 0$ for all ASQs and $E_{{\rm J}}/h = 10$~GHz.
    (d) Coupling strength between two ASQs, $i$ and $j$, versus the phase drop across one of them, $\phi_i$ (see Eq.~\ref{eq:coupling_N}) with the same parameters as in (c) and with $\phi_k=0$ for all other ASQs.
    The points on the $\phi_i$-axis with extremal and zero coupling strength are indicated with ON and OFF labels, respectively. 
    (e) Example of a design that minimizes the flux cross-coupling between loops.
    Each loop is implemented with a twisted gradiometric geometry that renders it insensitive to global magnetic field.
    The two subloops of each loop, shaded with the same color, have identical areas.
    The magnetic flux through each loop is controlled with a flux bias line, indicated with a hue-matching line.
    }
\end{figure*}

We propose a circuit that consists of a coupling junction, with Josephson energy $E_{\rm J}$ and phase drop $\phi$ across it, connected in parallel to a number, $N$, of ASQs, see Fig.~\ref{fig:scalability_concept}(a).
The subspace of each ASQ is spanned by two spin states, denoted as $\ket{\uparrow_i}$ and $\ket{\downarrow_i}$ for ASQ$i$, where $i$ is the qubit index.
This configuration defines $N$ loops through which magnetic fluxes, $\Phi_i$, are threaded, as indicated in Fig.~\ref{fig:scalability_concept}(a).

The individual ASQs are implemented using semiconducting quantum dot Josephson junctions~\cite{Lee2017, Fatemi2021, Kurilovich2021, Bargerbos2022, Canadas2022, Pavesic2024b, Sahu2024, Lakic2024}. 
The charging energy of each quantum dot is sufficiently large such that the ground state manifold is composed only of the two singly-occupied spin states.
Due to the spin-orbit coupling in the semiconductor, each ASQ is characterized by a spin-dependent, in addition to a spin-independent, Josephson energy, denoted as $E_{{\rm SO},i}$ and $E_{{\rm J},i}$, respectively~\cite{Padurariu2010, Bargerbos2023b}.
The values of $E_{{\rm SO},i}$ and $E_{{\rm J},i}$ can be tuned independently via electrostatic gates for each qubit (not shown in Fig.~\ref{fig:scalability_concept}).
The Hamiltonian of ASQ$i$ can be expressed in terms of these Josephson energies as
\begin{equation}
	H_i = E_{{\rm J},i} \sigma^0_i \cos\left(\phi_i\right) - E_{{\rm SO}, i} \sigma_i^z \sin\left(\phi_i\right)  + \frac{1}{2} \vec{E}_{{\rm Z},i} \cdot\vec{\sigma_i}\ , 
	\label{eq:scalability_ESOpotential}
\end{equation}
where $\sigma_i^z = \ket{\uparrow_i}\bra{\uparrow_i} - \ket{\downarrow_i}\bra{\downarrow_i}$ and $\sigma^0_i= \ket{\uparrow_i}\bra{\uparrow_i} + \ket{\downarrow_i}\bra{\downarrow_i}$ denote the $z$ Pauli operator and the identity operator of ASQ$i$, respectively, $\vec{\sigma_i}$ is the vector of $x$, $y$ and $z$ Pauli operators of ASQ$i$ and $\vec{E}_{{\rm Z},i}$ is the externally applied Zeeman field expressed in the coordinate system of ASQ$i$.

To lowest order in $E_{{\rm SO}, i}/E_{\rm J}$ and $E_{{\rm J}, i}/E_{\rm J}$, the phase drop through the coupling junction becomes $\phi=0$ and each $\phi_i$ is determined by the cumulative flux values from $1$ to $i$,
\begin{equation} \label{eq:scalability_phase}
  \phi_i = \sum_{j=1}^{i}{\varphi_j}, 
\end{equation}
where $\varphi_i = 2\pi\Phi_i/\Phi_0$, $\Phi_0=h/(2e)$ is the magnetic flux quantum, $h$ is the Planck constant and $e$ is the absolute value of the electron charge.
Therefore, by controlling the external fluxes, one can independently fix the values of all phase drops, $\phi_i$.
Taking the phase derivative of the ASQ Hamiltonian (Eq.~\ref{eq:scalability_ESOpotential}) and assuming a magnetic field aligned with the spin-polarization direction, $\vec{E}_{{\rm Z},i}\cdot\vec{\sigma_i}  = E_{{\rm Z},i} \sigma_i^z$, we obtain its current operator,
\begin{equation}\label{eq:scalability_current_operator}
  \mathcal{I}_i = \frac{\pi}{\Phi_0} \frac{\partial H_i}{\partial \phi_i} = \frac{I_{{\rm s}, i}}{2}\sigma^z_i + I_{0,i}\sigma^0_i,
\end{equation}
where $I_{0,i}=\frac{\pi}{\Phi_0}E_{{\rm J},i} \sin\left(\phi_i\right)$ represents the spin-independent component and the spin-dependent supercurrent is 
\begin{equation} \label{eq:scalability_supercurrent}
  I_{{\rm s}, i} = \frac{2\pi}{\Phi_0} E_{{\rm SO}, i}\cos\left(\phi_i\right).
\end{equation}
Notably, when $\phi_i$ is either $\pi/2$ or $-\pi/2$, $I_{{\rm s}, i}$ vanishes, rendering the supercurrent identical for both qubit states, see Fig.~\ref{fig:scalability_concept}(c).
On the contrary, when $\phi_i$ is either $0$ or $\pi$, the magnitude of the spin-dependent component of the supercurrent is maximal.
Throughout this manuscript, we refer to these two flux setpoints as OFF and ON, respectively as depicted in Fig.~\ref{fig:scalability_concept}(b).

\begin{figure*}
    \centering
    \includegraphics{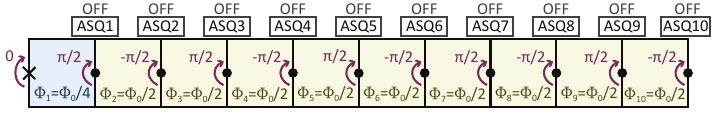} 
    \caption{\label{fig:scalability_idling}
        {\bf Idling flux configuration.} 
        Chain with $N=10$ showing the flux setpoint for an idling configuration in which all qubits are uncoupled. The choice of
        $\Phi_1 = \Phi_0/4$ and $\Phi_{i}=\Phi_0/2$ for all other loops sets alternating phase drops of $\pi/2$ and $-\pi/2$ for all qubits. 
  }
\end{figure*}

The circuit can be implemented in practice with independent control over the individual fluxes and with maximal addressability using the implementation illustrated in Fig.~\ref{fig:scalability_concept}(e).
Each loop is implemented with a twisted gradiometric loop geometry similar to the experimental implementation in Ref.~\cite{PitaVidal2024}.
Each loop is inductively coupled to a flux bias line with current $I_i$ and a symmetrical design at its end, where currents flow in opposite directions.
This combination of the loop and flux bias line designs maximizes their mutual inductance while minimizing unwanted cross-coupling to other loops.
Firstly, the two opposite currents on the flux line induce contributions to the flux through the loop that add up due to its twisted geometry.
Secondly, the generated magnetic field decreases with distance faster than for monopole flux line configurations,  
resulting in reduced magnetic fields at the locations of other loops.
Thirdly, the gradiometric loop design reduces the sensitivity to homogeneous fields, again reducing the unwanted cross-coupling, as well as the sensitivity to global magnetic noise.

We envision two possible driving mechanisms~\cite{Park2017, Cerrillo2021, Pavesic2024, Fauvel2024}: either using the flux bias lines for driving the spin-flip transitions via their supercurrent matrix element or applying microwave frequency pulses to the electrostatic gates of each qubit, which induces spin-transitions via the electric dipole spin resonance (EDSR) mechanism~\cite{Golovach2006, Nowack2007}.

The interactions between all pairs of qubits can be adjusted by varying the flux setpoints. 
For example, an idling configuration where all qubits are OFF is shown in Fig.~\ref{fig:scalability_idling}.
By setting $\Phi_1=\Phi_0/4$, ASQ1 is set to its OFF state, with $\phi_1=\pi/2$, i.e., the qubit does not interact with any other qubit. 
In turn, $\Phi_{i}=\Phi_0/2$ for the remaining loops results in alternating phase drops of $\pi/2$ and $-\pi/2$ for all other ASQs, rendering them OFF as well. 
Consequently, in this configuration all qubits are uncoupled. 

\section{All-to-all selective coupling}\label{s:scalability_coupling}

Next, we discuss the full circuit Hamiltonian and how we can control the interactions between multiple ASQs.
When either the Zeeman energy is low, $|\vec{E}_{{\rm Z},i}| \ll E_{{\rm SO},i}$, or the external magnetic field is applied along the spin-polarization direction for all qubits, $\vec{E}_{{\rm Z},i} = E_{{\rm Z},i} \sigma_i^z$, the qubits become pairwise longitudinally coupled to each other~\footnote{Note that sizable (compared to $E_{{\rm SO},i}$) perpendicular Zeeman components might introduce undesired transverse coupling terms between ASQs. 
See Appendix~\ref{ss:scalability_transverse}.}, as first discussed in Ref.~\cite{Padurariu2010} for the case of $N=2$ and experimentally realized in Ref.~\cite{PitaVidal2024}.
In this situation, the Hamiltonian describing the longitudinally coupled system can be expressed in the ASQ basis as
\begin{equation}
	H_{\rm ASQ} = \sum_{i=1}^{N} \left(  \frac{1}{2} E_i \sigma^z_i  + \sum_{j<i} \frac{1}{2} J_{ij}\sigma^z_i\sigma^z_j \right) , 
	\label{eq:scalability_H_longitudinal}
\end{equation}
where $E_i =  -2 E_{{\rm SO}, i}  \sin\left(\phi_i\right)  +  {E}_{{\rm Z},i, } $ is the energy of qubit $i$ and $J_{ij}$ represents the longitudinal coupling energy between qubits $i$ and $j$. In Eq.~\eqref{eq:scalability_H_longitudinal} and for the remainder of this section, we have disregarded spin-independent terms, as they have no influence on the spin dynamics.
Following Ref.~\cite{Padurariu2010} (see also Appendix~\ref{ss:scalability_higher_order}) the coupling strength is, to first order in $E_{{\rm SO}, i}/E_{\rm J}$, given by 
\begin{equation} \label{eq:coupling_N}
  J_{ij} = - 2\frac{E_{{\rm SO}, i}E_{{\rm SO}, j}}{|\tilde{E}|}   \cos{(\sum_{k=1}^{i}{\varphi_k}-\varphi_{\tilde{E}})}  \cos{(\sum_{k=1}^{j}{\varphi_k}-\varphi_{\tilde{E}})}, 
\end{equation}
where
\begin{equation}\label{eq:E_tilde_N}
\tilde{E} = E_{\rm J} + \sum_{l=1}^{N} E_{{\rm J}, l}e^{i  \sum_{k=1}^{l}{\varphi_k}} 
\end{equation}
is the total spin-independent Josephson energy of the system and $\varphi_{\tilde{E}}$ is the argument of $\tilde{E}$.

\begin{figure*}
    \centering
    \includegraphics{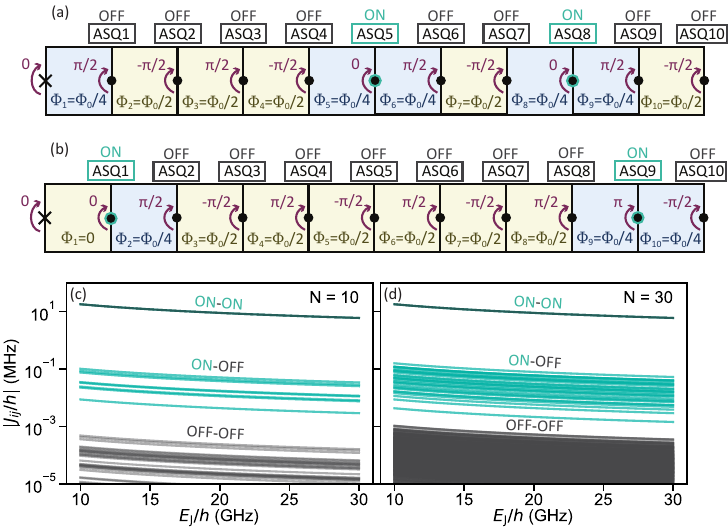} 
    \caption{\label{fig:scalability_coupling}
        {\bf All-to-all selective coupling.} 
        (a) and (b) Two chains with $N=10$ showing two example flux setpoints needed for selective two-qubit coupling between (a) qubits 5 and 8 and (b) qubits 1 and 9. 
        In contrast to the idling configuration of Fig.~\ref{fig:scalability_concept}(c), the phase drops across the two selected qubits, labeled as ON, are either $0$ or $\pi$, thus maximizing their spin-dependent supercurrent.
        In these configurations, the rest of the qubits remain uncoupled.
        (c) - (d) Absolute value of the qubit-qubit coupling strength, $|J_{ij}|$ obtained from Eq.~\ref{eq:coupling_N}, for two parameter configurations with random offsets added to the ideal flux bias points.
        For each panel, the dark green line indicates the coupling strengths between the two qubits that are ON, $n$ and $m$, the light green lines indicate the (undesired) coupling strengths between either $n$ or $m$ and another qubit and the grey lines indicate the (undesired) coupling strength between any other pair of qubits.
        (c) For $N=10$, $n=3$, $m=8$, $E_{{\rm SO}, i}/h = $~\SI{300}{MHz} and $E_{{\rm J}, i}/h = 0$ for all ASQs.
        The fluxes deviate from their ideal values by amounts $\Delta \Phi_i$ that take random values from a uniform distribution between plus and minus $0.001 \Phi_0$.
        (d) Same as (c) but for $N=30$, $n=6$ and $m=18$.         
  }
\end{figure*}

In the limit of $E_{{\rm J}, i}/E_{\rm J} \to 0$, the phase-offset $\varphi_{\tilde{E}}$ vanishes. 
In such scenario, it directly follows from Eq.~\ref{eq:coupling_N} that, when two qubits are ON (with $\sum_{k=1}^{i}{\varphi_k} = 0, \pi$), the coupling between them is maximal. 
Conversely, when either one of the two qubits is OFF (with $\sum_{k=1}^{i}{\varphi_k} = \pm \pi/2$), the coupling between them becomes zero, as illustrated in Fig.~\ref{fig:scalability_coupling}(a) and (b) for two possible flux configurations.

Away from the limit of $E_{{\rm J}, i}/E_{\rm J} \to 0$, the ON and OFF flux setpoints deviate from their exact values of $0, \pi$ and $\pm \pi/2$, respectively. The offset is $\varphi_{\tilde{E}}$, which depends on the global flux configuration and, importantly, can be independently measured.
Therefore, each individual flux can still be set to either maximize or turn off the couplings between any pair.
At this point it is interesting to note that when we adjust the flux configuration to the corrected flux setpoints, there is no unwanted coupling arising from the non-zero values of $E_{{\rm J}, i}$. The effect of $E_{{\rm J}, i}$, on the other hand, is to reduce the magnitude of the wanted coupling when their values become comparable to $E_{\rm J}$.

From Eq.~\ref{eq:coupling_N}, we calculate the coupling strength between any selected pair of qubits, see Fig.~\ref{fig:scalability_coupling}(c) and (d) for two examples with realistic parameter sets, with $N=10$ and $N=30$, respectively. We find ON-ON coupling strengths of around \SI{10}{MHz} that slowly decrease with increasing $E_{\rm J}$.
The couplings are calculated here for a situation in which two qubits, ASQ$n$ and ASQ$m$, are coupled to each other while the rest of the qubits are kept near their corrected OFF flux setpoints but deviate from them, each by a random value drawn from a uniform distribution between plus and minus $0.001 \Phi_0$. 
For typical experimental implementations, these deviations correspond to errors in the flux bias lines currents of less \SI{10}{\micro A}~\cite{PitaVidal2024, Wesdorp2024b}, well above the resolution of typical current sources~\cite{Yokogawa}, and thus well within experimental reach.
The unwanted ON-OFF and OFF-OFF couplings resulting from these imprecise flux settings remain significantly lower than the ON-ON coupling strength, by around 2 and 4 orders of magnitude, respectively, and can be reduced further by more precise flux control. 

By flux pulsing, this selective coupling scheme enables the implementation of CPHASE gates between any qubit pair within the system~\cite{Ma2024}.
In particular, coupling strengths of more than \SI{10}{MHz} would allow to realize CPHASE gates in less than $h/(4J)=\SI{25}{\nano\second}$.
Starting from an idling configuration, only the fluxes of the loops adjacent to the two selected qubits must be swept to reach the coupling configuration shown in Fig.~\ref{fig:scalability_coupling}(a) and (b).
Importantly, during pulsing, the two fluxes adjacent to qubit $m$ must be adjusted simultaneously to prevent undesired coupling between qubits $m$ and $m+1$.

\section{Quantum simulation of highly connected Ising systems}\label{s:scalability_simulation}

Beyond its use for digital gate-based quantum computation schemes, the system introduced in Fig.~\ref{fig:scalability_concept} holds potential for applications in analog quantum simulation.
The Hamiltonian presented in Eq.~\ref{eq:scalability_H_longitudinal}, which corresponds to either a reduced Zeeman field or a Zeeman field aligned with all qubits, directly maps the Hamiltonian of an all-to-all longitudinally connected Ising model.
More generally, when the Zeeman field has $E^x_{{\rm Z},i}\geq E_{{\rm SO}, i}$ components perpendicular to the spin directions, $\vec{E}_{{\rm Z},i}  = E^z_{{\rm Z},i} \sigma_i^z + E^x_{{\rm Z},i} \sigma_i^x$, the coupling Hamiltonian also includes transverse $\sigma_i^x\sigma_j^x$ terms (see Appendix~\ref{ss:scalability_transverse}):
\begin{equation} \label{eq:Hamiltonian_Ising}
  H_{\rm ASQ} = \sum_{i=1}^{N} \left(  \frac{E_i}{2} \sigma^z_i  + \sum_{j<i} \frac{J^{zz}_{ij}}{2} \sigma^z_i\sigma^z_j   + \sum_{j<i}\frac{J^{xx}_{ij}}{2} \sigma^x_i\sigma^x_j  \right).
\end{equation}
Here, $J^{zz}_{ij}$ and $J^{xx}_{ij}$ denote the longitudinal and transverse coupling energies, respectively.

\begin{figure*}
    \centering
    \includegraphics{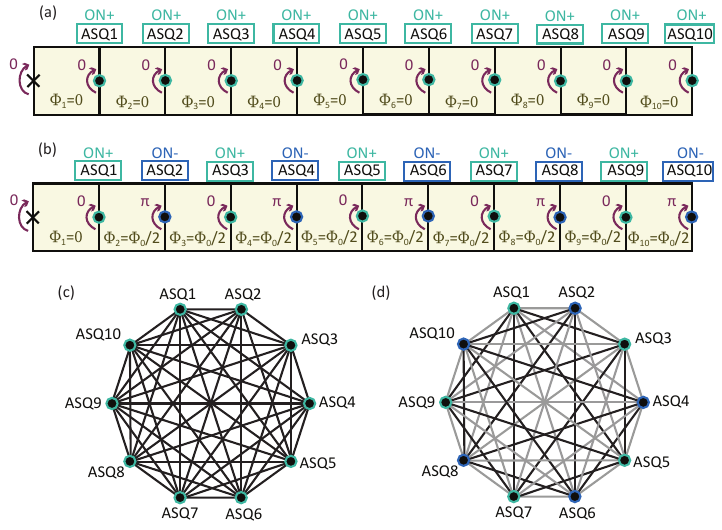} 
    \caption{\label{fig:scalability_all_to_all_coupling}
        {\bf Quantum simulation with ASQs.} 
        (a) and (b) Circuit diagrams for $N=10$ exemplifying two flux configurations in which all qubits are coupled to each other, thus mapping a highly connected Ising system. 
        (c) and (d) Graph diagrams indicating the couplings (edges) between each pair of qubits (nodes). 
        Panels (c) and (d) correspond to the flux configurations depicted in (a) and (b), respectively.
        For (c), all coupling strengths have equal sign  $J_{ij}=-|J_{ij}|$ (black edges). 
        For (d), the coupling strengths have either a negative sign, $J_{ij}=-|J_{ij}|$ if $|i-j|=2n$ (black edges), or a positive sign, $J_{ij}=+|J_{ij}|$ if $|i-j|=2n+1$ (grey edges), with $n$ being an integer number. 
  }
\end{figure*}

Classically, efficient simulation is possible only for sparsely connected longitudinal systems with planar couplings. 
However, when either transverse terms are present in a planar system or when the system exhibits higher connectivity, only nondeterministic polynomial time (NP)-hard classical exact solutions exist~\cite{Barahona1982}.
Flux qubit systems have been used to approach the solution of some problems using quantum annealing~\cite{Kadowaki1998, Somma2012, Albash2018}. 
Nonetheless, these quantum annealers have sparse connectivity, which requires an initial embedding of the desired problem into the qubit system at the cost of an increased number of physical qubits~\cite{Lechner2015}.
Due to their high connectivity, as we have introduced here, Andreev spin qubits constitute a promising solid-state platform well-suited for simulating a broader range of problems without requiring additional overhead in terms of qubits.
Besides applications in quantum annealing, this system can be used to explore the Ising spin dynamics without the need for Trotterization~\cite{Kim2023}.
The system presented here extends the range of Ising problems that can be simulated to encompass all partitioning problems, defined by $J^{zz}_{ij}=a^z_i a^z_j$ and $J^{xx}_{ij}=a^x_i a^x_j$.
However, a generic longitudinal Ising system has a total of $(N^2-N)/2$ independent couplings, meaning that the ASQ system studied here, with $N$ free flux parameters, cannot simulate all possible connectivity configurations of the Ising model.
Note that the tuning of the system from its fully uncoupled idling state (Fig.~\ref{fig:scalability_idling}) to a fully coupled state (Fig.~\ref{fig:scalability_all_to_all_coupling}(a) or (b)), only requires the adjustment of the flux through loop 1 by a quarter of a flux quantum. 
This provides straightforward control over the evolution time of an analog quantum simulation by only pulsing a single flux line.

\section{Readout}\label{s:scalability_readout}

The spin-supercurrent coupling of superconducting spin qubits provides a means for reading out their state through the use of circuit quantum electrodynamics techniques~\cite{Park2017, Hays2020, Metzger2021, Wesdorp2024, Bargerbos2023b}. 
In this section, we detail different readout alternatives that depend on the readout circuitry and on the specific qubits that need to be measured.
In Sec.~\ref{ss:sequential} we present a protocol for sequentially reading out the state of all qubits in the computational basis for a scenario in which the magnetic field is parallel to all qubits.
This can be achieved using either a transmon or a fluxonium circuit~\footnote{Note that, if the $E_{{\rm SO},i}/E_{\rm J}$ ratio needs to be reduced further by increasing $E_{\rm J}$ to the THz range, an alternative readout method involves replacing the coupling junction with a linear inductor and using a readout resonator with a geometry analogous to that in Refs.~\cite{Tosi2018, Hays2020, Metzger2021, Hays2021, Fatemi2021, Wesdorp2023, Canadas2022, Wesdorp2024, Sahu2024}.}. 
Subsequently, in Sec.~\ref{ss:selective}, we present a means to selectively read out the state of a single qubit while keeping all qubits uncoupled. 
Lastly, in Sec.~\ref{ss:scalability_joint_readout}, we instead present the joint readout of multiple qubits to determine the total number of qubits that are in their excited state.
Note that, in all cases, the readout circuit must be detuned from all ASQ frequencies to prevent transverse coupling between the qubits and the readout. 

\begin{figure*}
    \centering
    \includegraphics{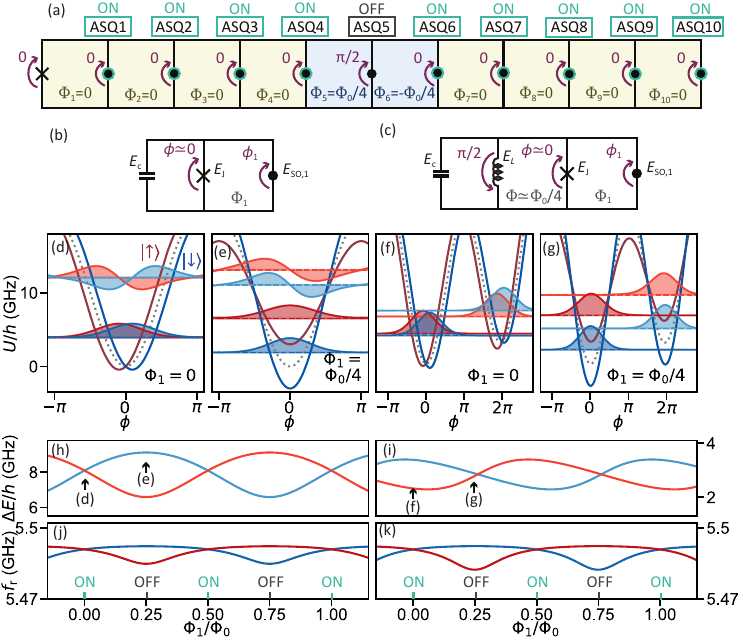} 
    \caption{\label{fig:scalability_readout}
        {\bf Sequential readout of qubits in their OFF setpoint.}    
        (a)  Circuit for reading out the state of qubit 5. 
        All qubits except for ASQ5 are ON. 
        The phase drop $\phi_5 \sim \pi/2$ sets ASQ5 OFF.
        (b) and (c) Two alternative readout circuits, shown for the case $N=1$ for simplicity. 
        (b) Transmon readout circuit diagram. 
        A capacitor with charging energy $E_{\rm c}$ is connected in parallel to the coupling junction and Andreev spin qubit.
        (c) Fluxonium readout circuit diagram, including also an inductor with inductive energy $E_L$ connected in parallel. 
        The inductor and the coupling junction define a loop with magnetic flux $\Phi$ through it.
        (d) and (e) Transmon potential versus the phase drop across the coupling junction, $\phi$, for $\Phi_1 =$~0 (ON) and $\Phi_1 =\Phi_0/4$ (OFF), respectively.
        In both cases, $E_{\rm J, 1}=0$, $E_{\rm c}/h=$~\SI{1.0}{GHz} and $E_{\rm J}/h=$~\SI{10.0}{GHz}. 
        The dotted lines indicate the case $E_{\rm SO, 1}=0$.
        The colored lines indicate the two possible potentials depending on the state of the ASQ, for $E_{\rm SO,1}/h=$~\SI{3}{GHz}.
        The color-filled regions represent the wavefunctions of the two lowest energy scales in each case, in arbitrary units.
        (f) and (g) Same as (d) and (e) but for the fluxonium circuit with $E_{L}/h=$~\SI{0.3}{GHz} and at $\Phi=\Phi_0/4$. 
        (h) First transmon transition frequency versus flux for the two possible spin states, $\ket{\uparrow}$ and $\ket{\downarrow}$, indicated with red and blue, respectively. 
        (j) Frequency of a readout resonator with bare frequency $f_{\rm r, 0}=$~\SI{5.5}{GHz} capacitively coupled to the transmon with a resonator-transmon coupling strength $g/h=$~\SI{200}{MHz}. 
        (i) First fluxonium transition versus flux for the two possible spin states. 
        (k) Same as (j) but for a readout resonator capacitively coupled to the fluxonium with resonator-fluxonium coupling strength $g/h=$~\SI{200}{MHz}. 
        }
\end{figure*}

\subsection{Sequential readout of all qubits}\label{ss:sequential}

The resonator and transmon circuits used to read out the spin in previous work are sensitive to the ASQ inductance~\cite{Hays2020, Hays2021, Wesdorp2024, Metzger2021, Bargerbos2023b, PitaVidal2023, PitaVidal2024}. 
As a result, these circuits are maximally sensitive to an ASQ state when the ASQ is in its OFF setpoint, and fully insensitive to it when the setpoint is ON.
In the absence of magnetic field or under the presence of a magnetic field parallel to all qubits, if all qubits are set to their ON setpoint, they are fully uncoupled from the readout circuit, but they are also maximally longitudinally coupled to each other.
In this case, their relative phases rotate over time, but the population in each computational basis state remains conserved as the coupling is longitudinal, thus preserving the measurement result. 
As illustrated in Fig.~\ref{fig:scalability_readout}, we can use this idea for reading out each of the qubits, by sequentially switching each ASQ to their OFF setpoint one by one (an example for qubit 5 is shown in Fig.~\ref{fig:scalability_readout}(a)) since an ASQ in its OFF setpoint couples strongly to the readout circuit. 
We discuss two alternative circuits for selectively reading out the state of a qubit when it is OFF while being insensitive to the states of the qubits that are ON.

\subsubsection{Transmon readout}

The first approach employs a 
transmon circuit~\cite{Koch2007, Pavesic2024, Gungordu2024}, as depicted in Fig.~\ref{fig:scalability_readout}(b), for simplicity for the case $N=1$.
The transmon consists of a capacitor, with charging energy $E_{\rm c}$, connected in parallel to the coupling and ASQ junctions. 
Its Hamiltonian can be expressed as
\begin{equation}
	H_{\rm t} = -4E_{\rm c}(\hat{n}-n_{\rm g})^2 + E_{\rm J}(1-\cos(\phi))  + H_{\rm ASQ}(\phi),
	\label{eq:scalability_transmon_ham}
\end{equation}
where $H_{\rm ASQ}(\phi)$ denotes the Hamiltonian of all ASQs in parallel, now including the spin-independent parts, which depends on all fluxes.
$\hat{n}$ is the conjugate charge of $\phi$, and $n_{\rm g}$ is the offset charge in the transmon island, expressed in units of the Cooper pair charge $2e$.

As shown in Fig.~\ref{fig:scalability_readout}(d), when an ASQ is in its ON setpoint ($\varphi_1=0$), the transmon's eigenstates have the same energy independently of the qubit state ($\ket{\uparrow}$ or $\ket{\downarrow}$). 
Consequently, the transmon transition frequencies are identical for both qubit states (see Fig.~\ref{fig:scalability_readout}(h)).
If the ASQ is instead in its OFF setpoint, the transmon eigenenergies change depending on the qubit state (red and blue in Fig.~\ref{fig:scalability_readout}(e))\cite{Bargerbos2023, PitaVidal2023}.
Fig.~\ref{fig:scalability_readout}(j) shows the resulting frequencies of a readout resonator capacitively coupled to the transmon with a coupling energy $g$.
The resonator frequencies are different when the qubit is in its OFF setpoint and identical when it is ON, thus allowing to selectively readout the state of individual OFF qubits. 
The difference between the two resonator frequencies is then the effective dispersive shift from the ASQ. For the parameters used in Fig.~\ref{fig:scalability_readout}, we find a maximal effective dispersive shift of 7.4~MHz.

\subsubsection{Fluxonium readout}

An alternative readout circuit to realize the same protocol is a
fluxonium circuit~\cite{Manucharyan2009}, shown in Fig.~\ref{fig:scalability_readout}(c). 
Its Hamiltonian can be expressed as
\begin{equation}
	H_{\rm f} = -4E_{\rm c}\hat{n}^2 + \frac{1}{2}E_L(\phi-\varphi)^2 + E_{\rm J}\left(1-\cos{ \left(\phi\right)}\right)  + H_{\rm ASQ}(\phi),
	\label{eq:scalability_fluxonium_ham}
\end{equation}
where $E_L$ is the inductive energy of the fluxonium shunting inductor and $\varphi=2\pi\Phi/\Phi_0$ denotes the reduced flux through the loop formed by the inductor and the coupling junction.
If the fluxonium flux is set to $\varphi=\pi/2$, the circuit can be used to selectively read out qubits in their OFF setpoint, analogously to the transmon case (see Fig.~\ref{fig:scalability_readout}(f, g, i, k)).

\begin{figure*}[ht!]
    \centering
    \includegraphics{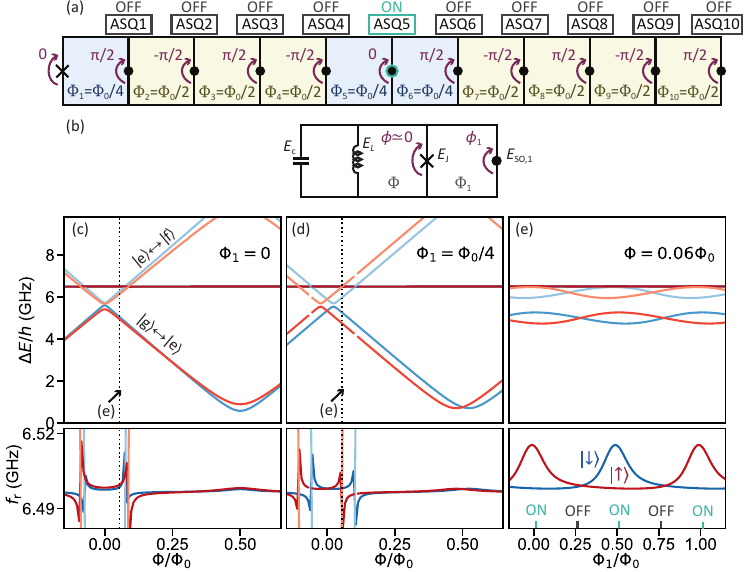} 
        \caption{\label{fig:scalability_readout_flux}
        {\bf Selective readout of a qubit in its ON setpoint.} 
        (a) Circuit diagram for $N=10$ exemplifying the flux configuration for reading out the state of qubit 5. 
        The phase drop $\phi_5 \sim 0$ sets ASQ5 ON, while all other qubits are OFF and thus uncoupled.
        (b) Same as Fig.~\ref{fig:scalability_readout}(c). In this case, the value of $\Phi$ is not fixed. 
        $E_{\rm c}/h=$~\SI{4.0}{GHz}, $E_{\rm SO,1}/h=$~\SI{1.5}{GHz}, $f_{\rm r, 0}=$~\SI{6.6}{GHz}, $g/h=$~\SI{300}{MHz} and the other parameters are the same as in Fig.~\ref{fig:scalability_readout}.
        (c), (d)  and (e) Show the frequencies of the fluxonium (top) and a resonator capacitively coupled to it (bottom) for different flux configurations.
        For (c) and (d) the ASQ is ON and OFF, respectively.
        As $\Phi$ is varied, there are different anticrossings between the resonator and the second fluxonium transition.
        (e) Shows the $\Phi_1$ dependence when $\Phi$ is fixed near one of such anticrossings, at the value indicated with a vertical dotted line in (c) and (d). 
        This results in equal resonator frequencies when the ASQ is OFF and in different resonator frequencies when it is ON. 
        }
\end{figure*}

\subsection{Selective readout of one qubit}\label{ss:selective}

In certain applications, such as ancilla-based parity readout, there is a need to selectively read out the state of an individual qubit without affecting any other qubits. 
To facilitate this selective readout, we introduce the protocol illustrated in Fig.~\ref{fig:scalability_readout_flux}.
This method consists of configuring all qubits to their OFF setpoints, ensuring they remain uncoupled. 
Simultaneously, the specific qubit that needs to be measured is set to its ON setpoint (see Fig.~\ref{fig:scalability_readout_flux}(a)). 
As only one qubit is ON, it does not interact with any of the other qubits.

As discussed in the previous section, a qubit that is ON can not be read out using the inductance-sensitive readout circuits presented in Fig.~\ref{fig:scalability_readout}.
Instead, we use a fluxonium circuit as the one shown in Fig.~\ref{fig:scalability_readout_flux}(b). 
This circuit is tuned to a precise flux $\varphi$ setpoint positioned near an avoided crossing between a higher-order fluxonium transition and the readout resonator. 
In fact, fluxonium qubits often get most of their dispersive shift from interactions with higher lying states~\cite{Zhu2013, Stefanski2023}. 
Fig.~\ref{fig:scalability_readout_flux}(c) and (d) show an avoided crossing between the second fluxonium transition and the readout resonator as $\varphi$ is varied, for the ON and OFF ASQ setpoints, respectively.
By setting $\varphi$ close to this avoided crossing, the frequency of the readout resonator depends on the ASQ state when the ASQ is ON and remains unaffected when it is OFF. 
This approach enables the selective readout of the state of an individual qubit in the ON setpoint while being insensitive to all other qubits that are in the OFF setpoint.
Note, however, that there is a balance between the proximity to the avoided crossing and the number of readout photons, as the closer the flux point is to the avoided crossing the higher the hybridization between the readout resonator and the fluxonium states~\cite{Nesterov2024}.

\subsection{Joint readout of all qubits}\label{ss:scalability_joint_readout}

In situations where the $E_{{\rm SO}, i}$ values of all qubits are similar, there is a third readout option available, which allows differentiation between various joint states based on the total number of qubits that are in their $\ket{\uparrow_i}$ state. 
This joint readout protocol entails configuring all qubits to their OFF flux setpoints, which leaves them uncoupled and thereby does not affect their state, and using the inductance-sensitive readout circuits introduced in Fig.~\ref{fig:scalability_readout}(b) and (c).
As a result, all joint states with the same number of qubits in their $\ket{\uparrow_i}$ state lead to the same dispersive shift on the resonator.
In total, there are $N$ different resonator frequencies, each corresponding to a different total number of $\ket{\uparrow_i}$ spins.
This configuration has several potential applications: (i) Direct counting of excited qubits. If the dispersive shifts and resonator line-width are designed so that the $N$ resulting readout signals can be distinguished, this configuration allows for the direct count of the total number of qubits in their $\ket{\uparrow_i}$ state. 
(ii) Measurement-induced state initialization. By selecting a specific readout frequency to distinguish the $\ket{\downarrow_0\downarrow_1 ... \downarrow_N}$ state from all other states, this approach can be employed for state preparation, to herald the system in this state~\cite{Johnson2012, Riste2012}. Such techniques can also be used to herald entangled states~\cite{Riste2013}.
(iii) Fidelity benchmarking of quantum gates. The ability to differentiate the $\ket{\downarrow_0\downarrow_1...\downarrow_N}$ state from all other states can be used to benchmark gate sequences that should have the global ground state as their final state, such as randomized benchmarking protocols~\cite{Magesan2011}.
(iv) Finally, in quantum simulation, distinguishing states with a fixed number of $\ket{\uparrow_i}$ spins from all other states can be useful to verify if the final state falls within the correct subspace, confirming the accuracy of the simulation \cite{Sagastizabal2019}.

These diverse applications underline the versatility of the ASQ system. However, further research is essential to address specific implementation details.

\section{Tune-up protocol}\label{s:scalability_tuneup}

\begin{figure}[h!]
	\centering
	\includegraphics{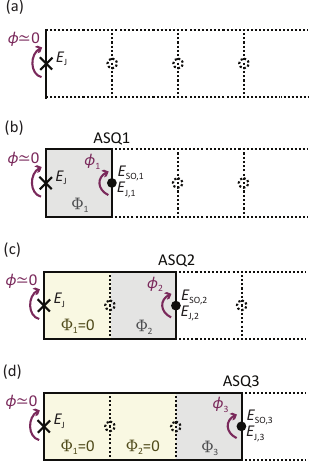} 
	\caption{\label{fig:scalability_tuneup}
      {\bf Sequential qubit tune-up.} 
      Each panel shows the circuit diagram of the loops array in different configurations, at subsequent steps in the tune-up process.
      (a) All qubits are pinched off with their electrostatic gates so that $E_{{\rm SO},i}=E_{{\rm J},i}=0$. 
      In this configuration, there are no loops and the value of $E_{\rm J}$ can be directly determined.
      (b) While keeping the rest of ASQs pinched off, ASQ1 is tuned up with its electrostatic gates following the procedure described in Ref.~\cite{Bargerbos2023b}.
      This allows to fix the desired values of $E_{{\rm SO},1}$ and $E_{{\rm J},1}$, as well as to determine the $I_1$ setpoints that set $\Phi_1=0, \Phi_0$.
      (c) Subsequently, the flux setpoint of ASQ1 is fixed at $\Phi_1=0$ and the electrostatic gates of ASQ1 are pinched off.
      Following a procedure analogous to that in (b) one can set the values of $E_{{\rm SO},2}$ and $E_{{\rm J},2}$ and determine the the $I_2$ setpoints that set $\Phi_2=0, \Phi_0$.
      (d) Same as (b) and (c) but for ASQ3.
  }
\end{figure}

To demonstrate the viability of this proposal, we now discuss a tuning protocol assuming the physical realization of ASQs as done in Refs.~\cite{PitaVidal2024, PitaVidal2023, Hays2021}.
Implementing each Andreev spin qubit in a semiconducting Josephson junction permits pinching it off (i.e. setting both of its Josephson energies to zero) by electrostatic gating.
In this section, we explain how, by selectively pinching off different combinations of qubits, the global system can be sequentially tuned up to its operational configuration, as depicted in Fig.~\ref{fig:scalability_tuneup}.

The tune-up process follows the procedure outlined in Ref.~\cite{Bargerbos2023b} and \cite{PitaVidal2024} for the $N=1$ and $N=2$ cases, respectively. 
In the initial step, illustrated in Fig.~\ref{fig:scalability_tuneup}(a), all ASQs are pinched off and the coupling junction is characterized.
As the $E_{\rm c}$ and $E_L$ values are known by design, the value of $E_{\rm J}$ can be determined from the measured frequency of the transmon or fluxonium readout circuitry.
If the coupling junction is implemented with a semiconducting Josephson junction, $E_{\rm J}$ can be electrostatically set at this step to a target value much larger than the target value for $E_{{\rm SO}, i}$ and $E_{{\rm J}, i}$. 

Subsequent steps involve selectively pinching off all ASQs except one, allowing it to be tuned up independently. 
This configuration is shown in Fig.~\ref{fig:scalability_tuneup}(b), (c) and (d) for qubits 1, 2 and 3, respectively.
This sequential approach permits the independent investigation of each qubit's gate and magnetic field dependences, enabling the selection of an optimal gate setpoint.

Initially, the gate space is mapped out at zero magnetic field to identify regions with sizable $E_{{\rm SO}, i}$ and low $E_{{\rm J}, i}$. 
This optimization aims at maximizing the coupling strength (see Eq.~\ref{eq:coupling_N}). 
An efficient way to perform this mapping is by detecting a frequency splitting of the readout resonator at fixed ASQ flux points (see Sec.~\ref{s:scalability_readout}).

Subsequently, the magnetic field is set to a non-zero value, allowing for the investigation of the magnetic field dependence. 
This step provides access to the spin-polarization direction and the $g$-factor at each gate setpoint.

During operation, the global magnetic field will be fixed at a predetermined value and direction, chosen to align with the chip plane to maximize the magnetic field resilience of the readout circuitry~\cite{Graaf2012, Samkharadze2016, Kroll2019, Zollitsch2019, FeldsteinIBofill2022}. 
Therefore, for each ASQ, a gate setpoint is selected so that the spin-polarization direction aligns with the preferred direction relative to the chosen magnetic field operation direction.
This relative alignment depends on the application.
For gate-based quantum computing, the operation is simplified if only longitudinal coupling terms are present (as in Eq.~\ref{eq:scalability_H_longitudinal}). 
Thus, the spin direction must either be chosen to be aligned with the magnetic field direction or, 
alternatively, the system can be operated under magnetic field strengths much lower than $E_{{\rm SO},i}$.
The latter option avoids the need for the spin polarization directions of all ASQs to be aligned with each other, thus simplifying the system tune up, and at the same time reduces the charge noise sensitivity~\cite{Lawrie2022}.
For quantum simulation of Ising systems, however, the spin direction can be adjusted to determine the ratio between longitudinal and transverse coupling terms. 
A consideration regarding the $g$-factors is that the qubit frequencies should not match the frequencies of the readout resonator or the readout superconducting qubit and that they lay within an experimentally accessible frequency band.
For InAs-based devices, typical $g$-factor values range from 2 to 16~\cite{Vaitekenas2018, Bargerbos2023b, Wesdorp2024}. 
This corresponds to frequencies between 1.5 and 11.2 GHz for an applied magnetic field of 50 mT.
Once the setpoint for one qubit is determined in this manner, its junction can be pinched off, and the next qubit can be characterized and tuned up similarly.

The change in the global magnetic field during the preceding steps alters the $\Phi_i(I_i)$ mappings. Therefore, these mappings can be determined in a subsequent round of tune-up steps, carried out after fixing the global magnetic field at its operational setpoint.
In this step, each qubit is sequentially opened (i.e. set to its gate setpoint) and an $I_i$ dependence is performed to determine the $I_i$ values that set $\Phi_i=0$ and $\Phi_i=\Phi_0$, as indicated in Fig.~\ref{fig:scalability_tuneup}.

\section{Discussion}\label{s:scalability_vision}

In the preceding sections, we discussed the potential of highly connected Andreev spin qubits for quantum computing and quantum simulation tasks. 
Our analysis focused on an estimation of the coupling strength to first order in $E_{{\rm SO},i}/E_{\rm J}$, as detailed in Sec.~\ref{s:scalability_coupling}. 
However, it is crucial to consider the contribution of the spin-dependent inductance of each ASQ, which becomes significant away from the limit $E_{{\rm SO},i}/E_{\rm J} \to 0$ (see Appendix~\ref{ss:scalability_higher_order}). 
These contributions introduce higher-order coupling terms of the form $\epsilon_{k} J_{ij} \sigma^z_i\sigma^z_j\sigma^z_k$, where $\epsilon_{k} \approx E_{{\rm SO},k}/E_{\rm J}$. 
For instance, the experimentally realistic values of $E_{{\rm SO},i}/h=$~\SI{300}{MHz} and $E_{\rm J}/h=$~\SI{30}{GHz} discussed in Fig.~\ref{fig:scalability_coupling} result in an ON-ON coupling strength of $J_{ij}/h =$~\SI{6}{MHz} and unwanted higher-order coupling terms between two ON qubits, $i$ and $j$, and an OFF qubit, $k$, of the order of $J_{ijk}/h =$~\SI{60}{kHz}.

To reduce the $E_{{\rm SO},i}/E_{\rm J}$ ratio and, consequently, mitigate the magnitude of higher-order terms, one can reduce the inductance of the coupling junction. 
A reduction to $E_{{\rm SO},i}/E_{\rm J} = 1/1000$ within experimentally attainable parameters can be realized by setting $E_{{\rm SO},i}/h=$~\SI{1}{GHz} and $E_{\rm J}/h=$~\SI{1000}{GHz}. 
This, in turn, sets the ON-ON coupling to $J_{ij}/h =$~2~MHz. 
Under these conditions, the higher-order coupling terms have an amplitude of $J_{ijk}/h =$~2~kHz.
This magnitude of $E_{\rm J}$ can be achieved by replacing the coupling junction with a linear inductor and using a resonator, instead of a transmon or fluxonium, for readout~\cite{Hays2020, Hays2021, Fatemi2021, Wesdorp2023, Wesdorp2024}. 
Note that the reduced coupling strength in this regime leads to slower dynamics and calls for qubit coherence higher than that of previous ASQ implementations \cite{PitaVidal2023, PitaVidal2024, Hays2021}.

\begin{figure}[!ht]
	\centering
	\includegraphics{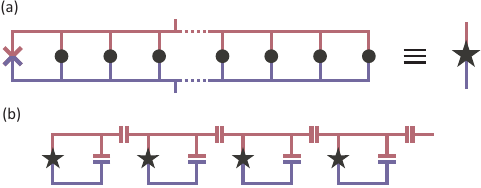} 
	\caption{\label{fig:scalability_vision}
      {\bf Quantum computing with superconducting spin qubits.} 
      (a) One cluster of $N$ coupled ASQs is represented with a star symbol.
      (b) Each cluster is read out with a transmon circuit.
      Different clusters are capacitively coupled to each other to mediate inter-cluster coupling.
  }
\end{figure}

These higher-order terms reduce the CPHASE gate fidelity, imposing a constraint, $N_{\rm max}$, on the maximum number of qubits in the system.
In particular, if $J_{ijk}/J_{ij} = \epsilon$ for all qubits,  we expect  a two-qubit gate infidelity of  $0.1875 ((N-2) \epsilon \pi)^2 $ 
\cite{Krinner2020}.
For $\epsilon = 0.001$, this results in $N_{\rm max, 99}=737$ 
and $N_{\rm max, 99.9}=234$ 
for target gate fidelities of 99.0~\% and 99.9~\%, respectively. 
Moreover, as discussed in Sec.~\ref{s:scalability_coupling}, imprecise flux control would lead to residual ON-OFF couplings, further reducing the gate fidelity.
If $J_{ijk}/J_{ij} = J_{ik}/J_{ij} = J_{jk}/J_{ij} = \epsilon$, the gate infidelity is $1.1875 ((N-2) \epsilon \pi)^2 $ which, for $\epsilon=0.001$, limits the number of qubits to $N_{\rm max, 99}=294$ or $N_{\rm max, 99.9}=94$ for the same target gate fidelities.

To scale up beyond this limit, we envision defining independent unit cells, each containing $N_{\rm max}$ qubits, as illustrated in Fig.~\ref{fig:scalability_vision}(a).
A potential architecture for coupling different clusters to each other involves capacitively coupling the readout circuit of two separate clusters (Fig.~\ref{fig:scalability_vision}(b)).
This hierarchical approach provides a scalable path for creating larger quantum processors while mitigating the impact of higher-order coupling terms.

To conclude, we have presented an approach for scaling up Andreev spin qubits in a highly connected way.
Our work demonstrates the ability to control the magnitude of the coupling strength between any pair of qubits, independently of their physical distance, by varying the applied flux. 
This feature enables the realization of fast two-qubit gates across the entire system.
Moreover, as all couplings can be made of the longitudinal type, the coupling strength remains independent of the relative qubit frequencies.
This characteristic offers great flexibility for increasing the qubit count without encountering issues associated with frequency crowding.

When it comes to gate-based digital quantum computation, this enhanced qubit connectivity opens up opportunities for exploring alternative quantum error correction codes, potentially requiring fewer physical qubits per logical qubit compared to existing surface codes~\cite{Bravyi2024}.
Regarding analog quantum simulation of Ising systems, the all-to-all connectivity extends the range of NP-hard problems that can be simulated in solid-state qubit platforms without the need for additional qubit overhead to encode the relevant problems, presenting an advantageous alternative to other superconducting qubit approaches.

\begin{acknowledgments}
We thank B. van Heck and V. Fatemi for discussions and their feedback on this manuscript. 
This work is co-funded by Microsoft Corporation.
C.K.A. acknowledges support from the Dutch Research Council (NWO) and from the Horizon Europe Framework Program of the European
Commission through the European Innovation Council
Pathfinder Grant No. 101115315 (QuKiT).
\end{acknowledgments}

\appendix

\section{Higher order longitudinal coupling terms}\label{ss:scalability_higher_order}

Away from the limit $E_{{\rm SO}, i}/E_{\rm J} \to 0$ it becomes essential to account for modifications to the coupling described in Eq.~\ref{eq:coupling_N} due to the presence of a state-dependent parallel inductance. This inductance introduces a modification to $\tilde{E}$, resulting in
\begin{equation}\label{eq:E_tilde_N_higher_order}
\tilde{E} = E_{\rm J} + \sum_{l=1}^{N} E_{{\rm J}, l}e^{i  \sum_{k=1}^{l}{\varphi_k}} + \sum_{l=1}^{N} \sigma^z_l E_{{\rm SO}, l}e^{i  (\frac{\pi}{2} +\sum_{k=1}^{l}{\varphi_k})}.
\end{equation}

The $E_{{\rm SO}, i}/E_{\rm J}$ term gives rise to higher order coupling terms of the form $\sigma^z_i \sigma^z_j\sigma^z_k$ that can be found by Taylor expanding Eq.~\ref{eq:coupling_N} around $E_{{\rm SO}, i}/E_{\rm J} = 0$.
The second-order contribution from the denominator yields
\begin{equation} \label{eq:coupling_N_higher_order}
  J^{(2)}_{ij} = J^{(1)}_{ij}  \left(1 -  \sum_{l=1}^{N} \frac{E_{{\rm SO}, l}}{E_{\rm J}} \sigma^z_l  e^{i  (\frac{\pi}{2} +\sum_{k=1}^{l}{\varphi_k})} \right),
\end{equation}
where $J^{(1)}_{ij}$ is the first-order approximation from Eq.~\ref{eq:coupling_N}.
This results in a contribution to the three-qubit coupling terms, $J_{ijk} \sigma^z_i \sigma^z_j\sigma^z_k $, that is a factor of $E_{\rm J}/E_{{\rm SO}, k}$ times smaller than the two-qubit coupling terms $J_{ij}$ (i.e. $J_{ijk} = E_{{\rm SO}, k}/E_{\rm J} J_{ij} \sigma^z_k $).

Additionally, Eq.~\ref{eq:E_tilde_N_higher_order} results in a $\sigma^z_k$-dependent contribution to $\varphi_{\tilde{E}}$.
This contribution can be as high as
\begin{equation} \label{eq:phase_error}
  \varphi_{\tilde{E}} = \tan^{-1} \left(\sum_{k=1}^{N}  \frac{E_{{\rm SO}, k}}{E_{\rm J}}\sigma^z_k    \right) \approx \sum_{k=1}^{N} \frac{E_{{\rm SO}, k} }{E_{\rm J}}\sigma^z_k 
\end{equation}
when all qubits are OFF. 
Here, the approximation is again made to first order in $E_{{\rm SO}, k}/E_{\rm J}$.
Similarly, this introduces a second contribution to the three-qubit coupling terms, given again by
\begin{equation} \label{eq:coupling_N_higher_order_phase}
  J_{ijk} = J_{ij}   \frac{E_{{\rm SO}, k} }{E_{\rm J}} \sigma^z_k.
\end{equation}
Once again, this term is scaled by a factor of $E_{{\rm SO}, k}/E_{\rm J}$ compared to the two-qubit coupling terms.

\section{Transverse coupling under the presence of a perpendicular Zeeman field}\label{ss:scalability_transverse}

For the sake of simplicity and without loss of generality, we consider the case $N=2$.
The Hamiltonian of the coupled ASQ system, expressed in the eigenbasis of the system at zero magnetic field and disregarding spin-independent terms, is given by
\begin{equation}
	H_{\rm ASQ} =  \frac{1}{2} E_1 \sigma^z_1 + \frac{1}{2} E_2 \sigma^z_2   + \frac{1}{2} J_{12}\sigma^z_1\sigma^z_2. 
	\label{eq:scalability_H_longitudinal_N2}
\end{equation}

If a perpendicular magnetic field is applied, the qubits eigenstates are no longer the same.
In particular, the Hamiltonian of ASQ$i$ becomes
\begin{align}
    H_i(\phi_i) & =   E_{{\rm J},i} \sigma^0_i \cos\left(\phi_i\right) - E_{{\rm SO}, i} \sigma_i^z \sin\left(\phi_i\right)  \\
                & \qquad  +  \frac{1}{2} E_{{\rm Z}, i} \cos{(\theta_i)} \sigma_i^z  + \frac{1}{2} E_{{\rm Z}, i} \sin{(\theta_i)} \sigma_i^x , 
    \label{eq:scalability_ESOpotential_theta}
\end{align}
where $E_{{\rm Z}, i}$ represents the magnitude of the Zeeman field and $\theta_i$ is the angle between the direction of the external Zeeman field and the zero-field spin direction of ASQ$i$.
In the limit of $E_{{\rm Z},i} \gg E_{{\rm SO},i}$, the eigenstates of ASQ$i$, expressed in its zero-field basis, become
\begin{align}
	\ket{\overline{\downarrow_i}} & = \left( \cos{(\frac{\theta_i}{2})}, \sin{(\frac{\theta_i}{2})} \right) {\rm and} \\
    \ket{\overline{\uparrow_i}} & = \left( -\sin{(\frac{\theta_i}{2})}, \cos{(\frac{\theta_i}{2})} \right).
\end{align}
Consequently, the zero-field $\sigma^z_i$ and $\sigma^x_i$, expressed in the new $\left\{\ket{\overline{\downarrow_i}}, \ket{\overline{\uparrow_i}}  \right\}$ basis, become
\begin{align}
	\sigma^z_i & = \cos{(\theta_i)} \sigma^{\overline{z}}_{i} + \sin{(\theta_i)} \sigma^{\overline{x}}_{i},  \\
    \sigma^x_i & = \cos{(\theta_i)}\sigma^{\overline{x}}_{i} + \sin{(\theta_i)} \sigma^{\overline{z}}_{i},
\end{align}
where $\sigma_i^{\overline{z}} = \ket{\overline{\uparrow_i}}\bra{\overline{\uparrow_i}} - \ket{\overline{\downarrow_i}}\bra{\overline{\downarrow_i}}$ and $\sigma_i^{\overline{x}} = \ket{\overline{\uparrow_i}}\bra{\overline{\downarrow_i}} + \ket{\overline{\downarrow_i}}\bra{\overline{\uparrow_i}}$.
This leads to the coupling Hamiltonian, expressed in the new spin eigenbasis
\begin{equation}
  \begin{split}
	H_{\rm ASQ}  & =  \frac{1}{2} E_1 \sigma^{\overline{z}}_{1} + \frac{1}{2} E_2 \sigma^{\overline{z}}_{2}  \\ 
               & + \frac{1}{2} J_{12} \cos(\theta_1) \cos(\theta_2) \sigma^{\overline{z}}_{1}\sigma^{\overline{z}}_{2} \\
               & + \frac{1}{2} J_{12} \sin(\theta_1) \cos(\theta_2) \sigma^{\overline{x}}_{1}\sigma^{\overline{z}}_{2} \\
               & + \frac{1}{2} J_{12} \cos(\theta_1) \sin(\theta_2) \sigma^{\overline{z}}_{1}\sigma^{\overline{x}}_{2} \\
               & + \frac{1}{2} J_{12} \sin(\theta_1) \sin(\theta_2) \sigma^{\overline{x}}_{1}\sigma^{\overline{x}}_{2}.
  \end{split}
	\label{eq:scalability_H_transverse_N2}
\end{equation}
This expression comprises both transversal ($XX$) and longitudinal ($ZZ$) coupling terms, along with $ZX$ and $XZ$ terms, with amplitudes
\begin{align}
	J_{12}^{zz} = & J_{12} \cos(\theta_1) \cos(\theta_2), \\
	J_{12}^{xz}  = & J_{12} \sin(\theta_1) \cos(\theta_2), \\
	J_{12}^{zx}  = & J_{12} \cos(\theta_1) \sin(\theta_2)  \hspace{0.14cm} {\rm and}\\
	J_{12}^{xx}  = & J_{12} \sin(\theta_1) \sin(\theta_2).
  \end{align}
If the Zeeman field is perpendicular to both qubits, $\theta_1=\theta_2=\pi/2$, the longitudinal, $XZ$ and $ZX$ terms vanish, leaving only the transverse coupling term with an amplitude of $J_{12}^{xx} = J_{12}$.

\vspace{1cm}

\bibliography{bibliography.bib}

\end{document}